\documentclass[english,aps,preprintnumbers,amsmath,amssymb,nofootinbib,twocolumn,10pt,prd]{revtex4-1} 

\usepackage[utf8]{inputenc}
\usepackage[T1]{fontenc}

\usepackage{ifthen}
\usepackage{forloop}
\usepackage{babel}
\usepackage{graphicx}
\usepackage{psfrag}
\usepackage{color}
\usepackage{dsfont}
\usepackage{mathrsfs}

\usepackage{slashed}

\usepackage{mathbbol}

\usepackage{ulem}

\usepackage{color}

\newcommand{\metric}{\ensuremath{g}}

\newcommand{\vielbein}{\ensuremath{e}}

\newcommand{\flatgamma}{{\ensuremath{\gamma_{\ttm{(\text{f})}}}}}
\newcommand{\tildeflatgamma}{{\ensuremath{\left. \tilde{\gamma}_{\ttm{(\text{f})}} \right. \! }}}

\newcounter{pointcounter}
\newcommand{\point}[1][\empty]{
  \ifthenelse
    {\equal{#1}{\empty}}
    {\ensuremath{\, \phantom{\cdot} \,}}
    {\setcounter{pointcounter}{1} \forloop{pointcounter}{0}{\value{pointcounter} < #1}{\ensuremath{\, \phantom{\cdot} \,}}}
}

\newcommand{\Eqref}[1]{Eq.~\eqref{#1}}
\newcommand{\ttm}[1]{\ensuremath{\text{\tiny{$#1$}}}}
\newcommand{\euler}{\mathrm{e}}
\newcommand{\cplx}{\mri}

\newcommand{\tr}{\operatorname{tr}}

\newcommand{\mcR}{\ensuremath{\mathcal{R}}}
\newcommand{\mcS}{\ensuremath{\mathcal{S}}}

\newcommand{\mri}{\ensuremath{\mathrm{i}}}
\newcommand{\mrI}{\ensuremath{\mathrm{I}}}

\newcommand{\C}{\ensuremath{\mathds{C}}}
\newcommand{\N}{\ensuremath{\mathds{N}}}
\newcommand{\R}{\ensuremath{\mathds{R}}}
\newcommand{\Z}{\ensuremath{\mathds{Z}}}

\makeatother

\begin{document}

\title{Global surpluses of spin-base invariant fermions} 
\author{Holger Gies}
\affiliation{Theoretisch-Physikalisches Institut, Friedrich-Schiller-Universit\"at Jena, 
Max-Wien-Platz 1, D-07743 Jena, Germany}                                        
\author{Stefan Lippoldt}
\affiliation{Theoretisch-Physikalisches Institut, Friedrich-Schiller-Universit\"at Jena, 
Max-Wien-Platz 1, D-07743 Jena, Germany}                                        

\begin{abstract}
The spin-base invariant formalism of Dirac fermions in curved space
maintains the essential symmetries of general covariance as well as
similarity transformations of the Clifford algebra. We emphasize the
advantages of the spin-base invariant formalism both from a conceptual
as well as from a practical viewpoint. This suggests that
local spin-base invariance should be added to the list of
(effective) properties of (quantum) gravity theories. We find support for
this viewpoint by the explicit construction of a global realization of
the Clifford algebra on a 2-sphere which is impossible in the
spin-base non-invariant vielbein formalism.
\end{abstract}

\maketitle

\section{Introduction}
The mutual interrelation of matter and spacetime (``matter curves
spacetime - spacetime determines the paths of matter'') is
particularly apparent for fermions. For instance for Dirac fermions,
information about both spin as well as spacetime meets in the Clifford algebra,
\begin{equation}
\{ \gamma_\mu,\gamma_\nu\} =2 \metric_{\mu\nu}
\mrI, \label{eq:CA}
\end{equation}
where the Dirac matrices $\gamma_\mu$ as well as the metric
$g_{\mu\nu}$ generally are spacetime dependent. While many tests of
classical gravity rely on vacuum solutions to Einstein's equation,
also many attempts at quantizing gravity primarily concentrate on the
dynamics of spacetime without matter,
cf. \cite{Ashtekar:2014kba}. This is similar in spirit to ``quenched''
QCD which allows to understand already many features of the strong
interactions at the quantum level even quantitatively.  Only recently,
some evidence has been collected that the existence of matter degrees
of freedom can constrain the existence of certain quantum gravity
theories
\cite{Percacci:2002ie,Eichhorn:2011pc,Dona:2012am,Dona:2013qba,Eichhorn:2015bna}.
This is again analogous to QCD where the presence of too many 
dynamical fermions can destroy the high-energy completeness of the
theory.

The interrelation of gravity and fermions provided by \eqref{eq:CA}
has also been interpreted in various partly conflicting directions:
read from right to left, one is tempted to conclude that one first
needs a spacetime metric $g_{\mu\nu}$ in order to give a meaning to
spinorial degrees of freedom and corresponding physical observables
such as currents $\sim\bar\psi\gamma_\mu\psi$. On the other hand,
representation theory of the Lorentz group in flat space suggests that
all nontrivial representations can be composed out of the fundamental
spinorial representation, culminating into \eqref{eq:CA} for Dirac
spinors. If so, then also the metric might be a composite degree of
freedom, potentially arising as an expectation value of composite
spinorial operators, see, e.g.,
\cite{Kraus:2002sa,Hebecker:2003iw,Diakonov:2011im}.

As a starting point to disentangle this hen-or-egg problem -- spinors
or metric first? -- we consider the Clifford algebra \eqref{eq:CA} as
fundamental in this work. We emphasize that this is different from a
conventional approach
\cite{Weyl:1929}%
, where one
starts from the analogous Clifford algebra in flat (tangential) space,
$ \{ {\gamma_\ttm{(\mathrm{f})}}_{a} , {\gamma_\ttm{(\mathrm{f})}}_{b}
\} = 2 \eta_{a b} \mrI$, with fixed ${\gamma_\ttm{(\mathrm{f})}}_{a}$
and then uplifts the Clifford algebra to curved space with the aid of
a \textit{vielbein} $\vielbein_{\mu}^{\point a} (x)$, such that
${\gamma_\ttm{(\vielbein)}}_{\mu} = \vielbein_{\mu}{}^{ a}
{\gamma_\ttm{(\mathrm{f})}}_{a}$ satisfies \eqref{eq:CA}. In addition
to diffeomorphism invariance, the vielbein approach supports a local
SO(3,1) symmetry of Lorentz transformations in tangential space,
i.e. with respect to the roman {\textit{bein}} index. By contrast, the
Clifford algebra \eqref{eq:CA} actually supports a bigger symmetry of
local similarity (spin-base) transformations in addition to general
covariance.

Developing a formalism that features this full spin-base invariance
has first been initiated by Schrödinger \cite{Schroedinger:1932} and
amended with the required spin metric by Bargmann \cite{Bargmann:1932}
in 1932. Surprisingly, it has been rarely used in the literature, see,
e.g.,
\cite{Kofink:1949,Brill:1957fx,Unruh:1974bw,Finster:1998ws,Casals:2012es,Gies:2013dca,Christiansen:2015},
or even reinvented \cite{Weldon:2000fr}. A full account of
the formalism also including spin torsion has recently been given in
\cite{Gies:2013noa}. Particular advantages are not only the inclusion
and generalization of the vielbein formalism. In a quantized setting,
it even justifies the widespread use of the vielbein as an auxiliary
quantity and not as a fundamental entity. Common quantization schemes
relying on the metric as fundamental degree of freedom remain
applicable also with fermionic matter. Hence, a Jacobian from the
variable transformation to the vielbein does not have to be accounted
for \cite{Gies:2013noa}.

In this work, we present further advantages of the spin-base invariant
formalism and discuss some general aspects in order to elucidate the
interplay between diffeomorphisms and spin-base transformations. We
point out various options of defining the spin-base group, differing
by the possible field content of further interactions and also
naturally permitting a $\mathrm{Spin}^c$ structure. Since the
conventional vielbein formalism can always be recovered within the
spin-base invariant formalism, it is tempting to think that the latter
is merely a technical perhaps overabundant generalization of the
former. We demonstrate that this is not the case by an explicit
construction of a global spin-base on the 2-sphere -- a structure
which is not possible in the conventional formalism because of global
obstructions from the Poincar\'e-Brouwer (hairy-ball) theorem. We
believe that this example is paradigmatic for the surpluses of the
spin-base invariant formalism.

\section{General covariance and spin-base invariance}
\label{sec:GCSPI}

Local symmetries are expected to be fundamental, since
symmetry-breaking perturbations typically contain relevant components
which inhibit symmetry emergence. Hence, we consider the local
symmetries of the Clifford algebra as fundamental. These are
diffeomorphisms (formalized by tensor calculus of the Greek indices)
and local similarity transformations of the Dirac matrices
\cite{Pauli:1936}, the spin base transformations,
\begin{equation}
\gamma_\mu \to \mcS \gamma_\mu \mcS^{-1}, \quad \psi \to \mcS \psi, \quad \bar\psi \to \bar\psi \mcS^{-1}.
\label{eq:spinbase}
\end{equation}
The $\gamma_\mu$ transformation leaves the Clifford algebra
\Eqref{eq:CA} invariant. The corresponding transformation of spinors
ensures that typical fermion bilinears and higher-order
interaction terms serving as building blocks for a relativistic
field-theory are also invariant, provided a suitable connection exists.
The latter should obey
\begin{equation}
\Gamma_\mu \to \mcS \Gamma_\mu \mcS^{-1} - (\partial_\mu \mcS)\mcS^{-1},
\label{eq:spinconnection}
\end{equation}
such that $\nabla_{\mu} =\partial_\mu+\Gamma_\mu$ forms a covariant
derivative with the standard covariance properties with respect to
both diffeomorphisms as well as spin base transformations. The
connection $\Gamma_\mu$ has explicitly been constructed in $d=4$
dimensions \cite{Weldon:2000fr,Gies:2013noa} as well as in lower
\cite{Lippoldt:2012} and higher dimensions \cite{Lippoldt:2015}. For
vanishing spin torsion \cite{Gies:2013noa}, the traceless part of
$\Gamma_\mu$ can fully be expressed in terms of the Dirac matrices and
their first derivatives (part of the terms can be summarized by
Christoffel symbols).

For simplicity, let us confine ourselves to the cases $d=4$ and $d=2$
(for generalizations, see \cite{Lippoldt:2015}). Here, the dimension of the irreducible
representation of the Clifford algebra is $d_\gamma=4$ and
$d_\gamma=2$, respectively. A natural choice for the group of spin
base transformations maintaining all invariance properties mentioned
above is then given by GL($d_\gamma,\C$).

However, GL($d_\gamma,\C$) contains continuous subgroups that act
trivially on the Clifford algebra. Considering the invariance
properties of the Clifford algebra as fundamental, trivial
subgroups appear redundant. Locally, elements of GL($d_\gamma,\C$) can
be decomposed into an SL($d_\gamma,\C$) element and two factors
proportional to the identity: a phase $\in$ U(1) and a modulus
$\in\R_+$. Confining ourselves to the nontrivial invariance
properties, hence suggest to identify the set of transformation
matrices $\mcS$ with the fundamental representation of
SL($d_\gamma,\C$). This special linear group still has redundancies as
its discrete center $\Z_{d_\gamma}$ does not transform the Dirac
matrices nontrivially.

The choice of the local spin-base group becomes only
relevant, once a dynamics is associated with the connection. For the
choice of SL($d_\gamma,\C$) and vanishing torsion, the corresponding
field strength $\Phi_{\mu \nu}$ satisfies the identity
\cite{Weldon:2000fr,Gies:2013noa}
\begin{equation}
\Phi_{\mu\nu} = [\nabla_\mu,\nabla_\nu]= \frac{1}{8} R_{\mu\nu\lambda\kappa} [\gamma^\lambda,\gamma^\kappa].
\label{eq:Phi}
\end{equation}
It is somewhat surprising as well as reassuring that -- out of the
large number of degrees of freedom in $\Gamma_\mu$ -- only those
acquire a nontrivial dynamics which can be summarized in the
Christoffel symbols and hence lead to the Riemann tensor on the
right-hand side of \Eqref{eq:Phi}. As a consequence, spin-base
invariance is also a (hidden) local symmetry of any special
relativistic fermionic theory in flat space with an automatically
trivial dynamics for the connection, even if kinetic terms of the
form $\sim \tr \gamma^\mu \Phi_{\mu\nu} \gamma^\nu$ ($\sim R$
Einstein-Hilbert) or $\sim \tr \Phi_{\mu\nu} \Phi^{\mu\nu}$ would be
added.

This is different if spin-base transformations are associated with
GL($d_\gamma,\C$). Then two additional abelian field strengths
corresponding to the U(1) and the non-compact $\R_+$ factors appear on
the right-hand side of \Eqref{eq:Phi} and thus introduce further
physical degrees of freedom. These correspond to the imaginary and
real part of the trace of the connection $\Gamma_\mu$.\footnote{The
  non-compact factor (real part of $\tr \Gamma_\mu$) can be removed by
  fixing the determinant of the spin metric, see
  \cite{Weldon:2000fr,Gies:2013noa}.}  Hence, the identification of
the spin base group is in principle an experimental question to be
addressed by verifying the interactions of fermions. In this sense,
one might speculate whether the hypercharge U(1) factor of the
standard model could be identified with the spin-base group provided
proper charge assignments are chosen for the different fermions. The
inclusion of the U(1) factor is particularly natural on manifolds that
do not permit a Spin structure (e.g., CP${}^2$) \cite{Lawson:1990}, as
it provides exactly for the necessary ingredient to define the more
general Spin$^c$ structure.


For the remainder of this work, it suffices to consider
SL($d_\gamma,\C$) as the group of spin-base transformations. Returning
to the hen-or-egg problem, \Eqref{eq:Phi} seems interpretable as
another manifestation of the intertwining of Dirac structure and
curvature, or spin-base and general covariance. However, a
clearer picture arises from an explicit coordinate transformation of
the Clifford algebra,
\begin{equation}
\{\gamma_\mu',\gamma_\nu'\} = 2g_{\mu\nu}' \mrI=2\frac{\partial x^\rho
}{\partial x'{}^\mu} \frac{\partial x^\lambda}{\partial x'{}^\nu} g_{\rho\lambda} \mrI = \!
\left\{ \! \frac{\partial x^\rho
}{\partial x'{}^\mu} \gamma_\rho, \frac{\partial x^\lambda}{\partial x'{}^\nu} \gamma_\lambda \! \right\} \! .
\label{eq:CAdiff}
\end{equation}
Read together with the spin-base invariance of the Clifford algebra
\cite{Pauli:1936,Cornwell:1989bx}, \Eqref{eq:CAdiff} implies that the
most general coordinate transformation of a Dirac matrix is given by
\begin{equation}
\gamma_\mu\to\gamma_\mu' = \frac{\partial x^\rho}{\partial x'{}^\mu} 
\mcS\gamma_\rho \mcS^{-1}.
\label{eq:gengamtrafo}
\end{equation}
From the sheer size of the spin-base group (at least
SL($d_\gamma,\C$)), it is obvious that this is a larger set of Dirac
matrices satisfying the Clifford algebra than can be spanned by the
vielbein construction. In the latter, only those realizations of the
Clifford algebra ${\gamma_\ttm{(\vielbein)}}_{\mu}$ are considered,
that can be spanned by a fixed set of Dirac matrices,
${\gamma_\ttm{(\vielbein)}}_{\mu} = \vielbein_{\mu}{}^{ a}
{\gamma_\ttm{(\mathrm{f})}}_{a}$. A local Lorentz transformation with
respect to the \textit{bein} index can then be rewritten in terms of
\begin{equation}
\Lambda_{a}{}^{b} \flatgamma_b=\mcS_{\text{Lor}} \flatgamma_a \mcS_{\text{Lor}}^{-1},
\label{eq:Lor}
\end{equation}
where $\mcS_{\text{Lor}}\in \text{Spin}(d-1,1) \subset
\text{SL}(d_\gamma,\C)$. Conventionally, the $\mcS_{\text{Lor}}$
factors are interpreted as Lorentz transformations of Dirac spinors,
e.g., $\psi \to \mcS_{\text{Lor}} \psi$. This way of interpreting
the Lorentz subgroup of spin-base transformations is at the heart of
understanding fields as representations of the Lorentz group. This
viewpoint is held to argue that higher-spin fields (such as the
metric) may eventually be composed out of a fundamental spinorial
representation.

However, there is no such simple relation as \Eqref{eq:Lor} for
general coordinate transformations. This is already obvious in flat
space: rescaling one coordinate axis, say, $x^3\to x^3/\alpha$,
implies a change of the metric,
\begin{equation}
g_{\mu\nu}=\text{diag}(-1,1,1,1) \to g'_{\mu\nu} = \text{diag}(-1,1,1,\alpha^2).
\label{eq:rescaling}
\end{equation}
The corresponding change of the Dirac matrices cannot be written
purely in terms of a spin-base transformation, since $\gamma'_3$ has
to satisfy $(\gamma'_3)^2 = \alpha^2 \mrI$, whereas
$(\mcS\gamma_3\mcS^{-1})^2 = \mrI$ for all $\mcS\in$ SL($d_\gamma,\C$).

It is therefore more natural to view the general coordinate
transformation \eqref{eq:gengamtrafo} of the Dirac matrices as
consisting of two independent transformations, \text{(i)} the change
of the spacetime basis (diffeomorphisms), $\gamma_\mu\to
\frac{\partial x^\rho}{\partial x'{}^\mu} \gamma_\rho$, and
\text{(ii)} the change of the spin base \Eqref{eq:spinbase}. In
particular, there is no need to intertwine these
transformations.

While these statements seem self-evident within the present
discussion, they may appear uncommon if compared to the conventional
reasoning in flat space. Coordinate transformations between two
different Lorentz frames, $\Lambda_{a}{}^b= \frac{\partial
  x^b}{\partial x'{}^{a}}$, are typically combined with spin-base
transformations using \Eqref{eq:Lor} in order to keep the Dirac
matrices in the new frame form-identical to those in the old frame,
$\flatgamma'_a \equiv\flatgamma_a$. This is, however, not necessary,
as also $\gamma'_{a} = \Lambda_a{}^b \flatgamma_b$ satisfies the 
Clifford algebra. 

To summarize, spinors should be viewed as objects that transform as
scalars under diffeomorphisms and as ``vectors'' under spin-base
transformations. In flat space, Lorentz transformations and spin-base
transformations may be combined in order to keep the Dirac matrices
fixed. We emphasize that the latter is merely a convenient choice and
by no means mandatory. In fact, the freedom not to link the two
transformations can have significant advantages as shown in the next
section.

In view of the hen-or-egg problem, this symmetry analysis does not
single out a specific viewpoint. On the one hand,
the representation theory of the Lorentz group suggesting ``spinors
first'' should be embedded into the larger spin-base invariant
framework; while this presumably does not change the result for the
classification of fields, there is no analogue of \Eqref{eq:Lor} for
general spin-base transformations. On the other hand, the fact that we
need a metric to define the Clifford algebra, does not link spinors
closer to the metric as other fields; diffeomorphisms leave spinors
untouched and the transformed Dirac matrices satisfy the Clifford
algebra automatically. Instead, our analysis rather suggests that not
only local Lorentz invariance, but full local spin-base invariance
should be a requirement for possible underlying quantum theories of
gravity. If not at the fundamental level, local spin-base invariance
should at least be emergent for the long-range effective description.

\section{Global spin base}\label{subsec:relation_to_vielbein_formalism}

It is a legitimate question as to whether spin-base invariance
introduces an overabundant symmetry structure without gaining any
advantages or further insights. In fact, already the vielbein
formalism with much less symmetry has been criticized for its
redundancy. For instance, the Ogievetsky-Polubarinov spinors \cite{Ogievetsky:1965ii}
not only remove the SO(3,1) redundancy of the vielbein formulation
(analogous to the Lorentz symmetric gauge for the vielbein
\cite{Woodard:1984sj}), but make spinors compatible with tensor
calculus, see, e.g., \cite{Pitts:2011jv}.

Nevertheless, SL($d_\gamma,\C$) spin-base invariance is not a symmetry
that may or may not be constructed on top of existing symmetries. On
the contrary, global spin-base invariance is present in any
relativistic fermionic theory. Its local version does not need an
additional new compensator field, but the connection $\Gamma_\mu$ is
built from the Dirac matrices which are present anyway. We will now
present an example which demonstrates the advantages of full spin-base
invariance.

Rather generically, smooth orientable manifolds may not be
parametrizable with a single coordinate system, but may require
several overlapping coordinate patches. In the vielbein formalism,
where $g_{\mu\nu}= \vielbein_\mu{}^a\eta_{ab} \vielbein_\nu{}^b$, it
is natural to expect that patches with different coordinates and
corresponding metrics $g_{\mu\nu}$ also require different vielbeins
$\vielbein_\mu{}^a$. This becomes most obvious for the simple example
of a 2-sphere which requires at least two overlapping coordinate
patches to be covered. The same is true for the vielbein: for each
fixed bein index, $\vielbein_\mu{}^a$ is a spacetime vector which has
to satisfy the Poincar\'e-Brouwer (hairy-ball) theorem. This implies that
it has to vanish at least on one point of the 2-sphere (such that also
$\det \vielbein=0$). Hence, at least two sets of vielbeins and
corresponding transition functions are required to cover the 2-sphere
without singularities.

For the spin-base invariant formalism, the independence of
diffeomorphisms and spin-base transformations suggests that a change
of the coordinate patch and metric does not necessarily require a
change of the spin-base patch. More constructively, two sets of spin
bases on two neighboring coordinate patches may be smoothly connected
by a suitable spin-base transformation. We now show that this is
possible for the 2-sphere resulting in a global spin base.

To keep this discussion transparent, we use the pair of polar and
azimuthal angles $(\theta, \phi)$ to label all points on the sphere
(not as coordinates).  For the polar coordinates, we use the notation
$(\vartheta,\varphi)$, i.e., $(x^{\mu})|_{(\theta, \phi)} =
(\vartheta, \varphi)|_{(\theta, \phi)} = (\theta, \phi)$. These
are legitimate coordinates except for the poles at $\theta \in \{0, \pi
\}$. In polar coordinates the metric reads
\begin{align}
 \big(\metric_{\mu \nu}|_{(\theta, \phi)} \big) = \left. \begin{pmatrix} 1 & 0 \\ 0 & \sin^2 \vartheta \end{pmatrix} \right|_{(\theta, \phi)} = \begin{pmatrix} 1 & 0 \\ 0 & \sin^2 \theta\end{pmatrix} \! \text{.}
\end{align}
Obviously, the metric becomes degenerate at the poles, rendering the coordinates ill-defined there. In these coordinates, one suitable choice for the vielbein $\vielbein_{\mu}^{\point a}$ is
\begin{align}\label{eq:def:vielbein_on_sphere}
 \big( \vielbein_{\mu}^{\point a}|_{(\theta, \phi)} \big) = \left. \begin{pmatrix} 1 & 0 \\ 0 & \sin \vartheta \end{pmatrix} \right|_{(\theta, \phi)} = \begin{pmatrix} 1 & 0 \\ 0 & \sin \theta \end{pmatrix} \! \text{.}
\end{align}
This choice is perfectly smooth everywhere, but is not appropriate at the poles.
In order to cover the poles $\theta \in \{ 0 , \pi \}$, we need to change coordinates.
For simplicity, we choose Cartesian coordinates $(x'^{\mu})|_{(\theta, \phi)} = (x,y)|_{(\theta, \phi)} = (\cos \phi \sin \theta, \sin \phi \sin \theta)$, these are well defined at the poles but ill defined at the equator.
For the coordinate transformation, we need the Jacobian
\begin{align}
 \left( \left. \frac{\partial x^{\nu}}{\partial x'^{\mu}} \right|_{(\theta, \phi)} \right) 
 = \begin{pmatrix} \frac{\cos \phi}{\cos \theta} & - \frac{\sin \phi}{\sin \theta} \\ \frac{\sin \phi}{\cos \theta} & \frac{\cos \phi}{\sin \theta} \end{pmatrix} \! \text{.}
\end{align}
We emphasize again that the pair $(\theta, \phi)$ is used only for
convenience to label a point on the sphere and not as a coordinate system.
The metric for the primed (Cartesian) coordinates $x'{}^\mu$ reads
\begin{align}
 \big( \metric'_{\mu \nu}|_{(\theta , \phi)} \big) =& \frac{1}{1 - x^2 - y^2} \left. \begin{pmatrix} 1 - y^2 & x y \\ x y & 1 - x^2 \end{pmatrix} \right|_{(\theta, \phi)}
\\
 =& \frac{1}{\cos^{2} \theta} \begin{pmatrix} 1 - \sin^{2} \phi \sin^{2} \theta & \sin \phi \cos \phi \sin^{2} \theta \\ \sin \phi \cos \phi \sin^{2} \theta & 1 - \cos^{2} \phi \sin^{2} \theta \end{pmatrix} \! \text{.}
\end{align}
The transformed vielbein $\vielbein'{}_{\mu}^{\point a}$ yields
\begin{align}
 &\big( {\vielbein'}_{\mu}^{\point a}|_{(\theta , \phi)} \big) = \left( \left. \frac{\partial x^{\nu}}{\partial x'^{\mu}} \vielbein_{\nu}^{\point a} \right|_{(\theta , \phi)} \right) = \frac{1}{\cos \theta} \mcR_{\phi} \begin{pmatrix} 1 & 0 \\ 0 & \cos \theta \end{pmatrix} \! \text{,}
\\
 &\mcR_{\phi} = \begin{pmatrix} \cos \phi & - \sin \phi \\ \sin \phi & \cos \phi \end{pmatrix} \! \text{.}
\end{align}
First, we observe a coordinate singularity at the equator as
expected. Moreover, we obtain a $\phi$ dependence which seems to
render the vielbein ill defined at the poles. Nevertheless, this can
be cured by performing a corresponding (counter-)rotation in
tangential space with respect to the bein index. The hairy ball
theorem manifests itself here by the fact that one pole needs a
rotation, while the other needs a combination of the same rotation and
a reflection. These are elements of the two different connected
components of the rotation group $\mathrm{O}(2)$, respectively, the
proper and improper rotations. Since we cannot perform a continuous
transformation from proper to improper rotations, we cannot cure the
residual $\phi$ dependence at both poles at the same time in a
continuous way (independently of the expected coordinate singularity at
the equator). Incidentally, an inverse rotation would also cure the
problematic $\phi$ dependence at the south pole; but because of the
required $2\pi$ periodicity in $\phi$, the direction of the rotation 
cannot be changed continuously from north to south pole. The same
conclusion remains true for those sets of Dirac matrices which are
constructed via the vielbein ${\gamma'_{\ttm{(\vielbein)}}}_{\mu} |_{(\theta
  , \phi)} = {\vielbein'}_{\mu}^{\point a}|_{(\theta, \phi)}
\flatgamma_{a}$.

By contrast, the spin-base invariant formalism allows to continuously
connect all representations of the two dimensional Clifford algebra,
i.e., proper and improper rotations of O(2) should be continuously
connectable on the level of SL(2,\C) spin-base transformations of the
Dirac matrices.

For this, we first define conventional \textit{constant} flat Dirac matrices using the Pauli matrices
\begin{align}\label{eq:def:flatgamma_on_sphere}
 (\tildeflatgamma_{a}) = \begin{pmatrix} \sigma_{1} \\ - \sigma_{2} \end{pmatrix} \! ,
\end{align}
which fulfill the two dimensional flat Euclidean Clifford algebra $\{
\tildeflatgamma_{a} , \tildeflatgamma_{b} \} = 2 \delta_{ab} \mrI$.
Next, we construct auxiliary spacetime \textit{dependent} flat Dirac
matrices,
\begin{align}
 \flatgamma_{a}|_{(\theta , \phi)} &= \mcS(\theta, \phi) \tildeflatgamma_{a} \mcS^{-1}(\theta, \phi),
\label{eq:auxgam}
 \\
 \mcS(\theta , \phi) &= \euler^{- \cplx \frac{\phi}{2} \sigma_{3} } \euler^{- \cplx \frac{\theta - \pi}{2} \sigma_{1} }
 \text{,}
\label{eq:sbtrafo}
\end{align}
which also satisfy the Euclidean Clifford algebra as \Eqref{eq:auxgam}
is a spin-base transformation. We emphasize that \Eqref{eq:sbtrafo}
goes beyond the subgroup of $\mathrm{Spin}(2)$ transformations because
of the second exponential factor.
The new flat Dirac matrices read explicitly
\begin{align}
 \flatgamma_{1}|_{(\theta , \phi)} &= \cos \phi \sigma_{1} + \sin \phi \sigma_{2},
\\
 \flatgamma_{2}|_{(\theta , \phi)} &= \cos \theta ( - \sin \phi \sigma_{1} + \cos \phi \sigma_{2} ) + \sin \theta \sigma_{3} \text{.}
\end{align}
Here it becomes manifest, that these Dirac matrices smoothly vary from a proper rotation
at $\theta = 0$ to an improper rotation at $\theta = \pi$, while maintaining $2\pi$-periodicity in $\phi$. 
Based on this special set of flat-space Dirac matrices, the Dirac matrices on the 2-sphere in Cartesian coordinates
$\gamma'_\mu=\vielbein'{}_{\mu}{}^a \flatgamma_{a}$ read
\begin{align}\label{eq:curved_gamma_cartesian_coordinates}
 \big( \gamma'_{\mu}|_{(\theta , \phi)} \big) \! =& \frac{1}{\cos \theta} \mcR_{\phi} \! \begin{pmatrix} 1 \! & 0  \\ 0 \! & \cos^{2} \theta \end{pmatrix} \!\! \mcR_{\phi}^{-1} \! \begin{pmatrix} \sigma_{1} \\ \sigma_{2} \end{pmatrix} \!\!
 + \sin \theta \mcR_{\phi} \! \begin{pmatrix} 0 \\ \sigma_{3} \end{pmatrix} \! \text{.}
\end{align}
These Dirac matrices are obviously well behaved at the poles $\theta
\in \{ 0 , \pi \}$, since there are no singularities and no $\phi$
dependence is left. Of course, the singularity at the equator remains,
where the Cartesian coordinates are ill defined. This singularity is
not present in polar coordinates, where we obtain the Dirac matrices
$\gamma_\mu= \frac{\partial x'{}^{\rho}}{\partial x_\mu} \gamma'_\rho$
\begin{align}\label{eq:curved_gamma_polar_coordinates}
 \big( \gamma_{\mu}|_{(\theta , \phi)} \big) = \begin{pmatrix} 1 & 0 \\ 0 & \frac{1}{2}\sin 2 \theta \end{pmatrix} \mcR^{-1}_{\phi} \begin{pmatrix} \sigma_{1} \\ \sigma_{2} \end{pmatrix} + \sin^{2} \theta \begin{pmatrix} 0 \\ \sigma_{3} \end{pmatrix}\!\! \text{.}
\end{align}
Note that $\gamma_\mu$ and $\gamma'_\mu$ are connected solely by a
diffeomorphism -- no change of the spin base is involved. Whereas the
vielbein construction given above actually proceeded via ill-defined
intermediate objects\footnote{With hindsight, the ill-definiteness of
  ${\gamma_{(\vielbein)}}_{\mu}|_{(\theta , \phi)}$ at the poles is
  cured by the properties of the flat gamma matrix
  $\flatgamma_{a}|_{(\theta , \phi)}$ which are analogously ill
  defined at the poles.}, the resulting spin base chosen for the
curved Dirac matrices given by
\Eqref{eq:curved_gamma_cartesian_coordinates} in Cartesian coordinates
(i.e.~except for the equator) and by
\Eqref{eq:curved_gamma_polar_coordinates} in polar coordinates
(i.e.~except for the poles) holds globally all over the 2-sphere. No
additional patch for spin-base coordinates is required to cover the
whole 2-sphere. In particular the limit towards the poles in
\Eqref{eq:curved_gamma_cartesian_coordinates} is unique and smooth in
contrast to the vielbein case.

It is interesting to see how the spin-base invariant formalism evades
the hairy-ball theorem: the important point is that $\gamma_\mu$ does
not represent a globally non-vanishing vector field (which would be
forbidden), but is a vector of Dirac matrix fields,
$(\gamma_{\mu})^{I}_{\point J}$. For every fixed pair $(I,J)\in \{1,
\ldots , d_{\gamma} \}^{2}$, we have a complex vector field. It is
easy to check that each of the real sub-component vector fields has at
least one zero on the sphere, being therefore compatible with the
hairy-ball theorem. The zeros of these vector fields are however
distributed such that the Dirac matrices $\gamma_\mu$ satisfy the
Clifford algebra all over the 2-sphere.

We expect that the construction above generalizes to all $2n$-spheres,
since the corresponding spin-base group SL$(d_{\gamma},\C)$ with
$d_\gamma=2^n$ is connected and all representations of the Dirac
matrices are connected to each other via a spin-base
transformation. The problem of the disconnected components of the
orthogonal group should then be resolvable in the same way as shown
above. Incidentally, the hairy-ball theorem applies to the
$2n$-spheres, implying that vielbeins cannot be defined globally on
these spheres.

As an application of this global spin base, let us study the
eigenfunctions of the Dirac operator on the 2-sphere. Using the
vielbein $\vielbein_{\mu}^{\point a}$ of
\Eqref{eq:def:vielbein_on_sphere} and the flat Dirac matrices
$\tildeflatgamma_{a}$ of \Eqref{eq:def:flatgamma_on_sphere}
the eigenfunctions have been calculated in \cite{Camporesi:1995fb} within
the vielbein formalism. The vielbein spin connection
${\Gamma_{(\vielbein)}}_{\mu} = \frac{1}{8} \omega_{\mu}^{\point a b}
[ \tildeflatgamma_{a} , \tildeflatgamma_{b} ]$ in spherical
coordinates is then given by
\begin{align}
 \big( {\Gamma_{(\vielbein)}}_{\mu}|_{(\theta , \phi)} \big) = \begin{pmatrix} 0 \\ \frac{\cplx}{2} \cos \theta \sigma_{3} \end{pmatrix} \! \text{.}
\end{align}
The eigenfunctions of the Dirac operator $\slashed{\nabla} = {\gamma_{(\vielbein)}}^{\mu} (\partial_{\mu} + {\Gamma_{(\vielbein)}}_{\mu})$ satisfy 
$ \slashed{\nabla} \psi^{(s)}_{\pm, n, l} = \pm \cplx (n + 1) \psi^{(s)}_{\pm, n, l}$, $s \in \{ -1, 1 \}$ and read \cite{Camporesi:1995fb}
\begin{align}
 \psi^{(-)}_{\pm , n , l}(\theta , \phi) &= \frac{c_{2}(n,l)}{\sqrt{4 \pi}} \euler^{- \cplx (l + \frac{1}{2}) \phi} \begin{pmatrix} \Phi_{n,l}(\theta) \\ \pm \cplx \Psi_{n,l}(\theta) \end{pmatrix},
\\
 \psi^{(+)}_{\pm , n , l}(\theta , \phi) &= \frac{c_{2}(n,l)}{\sqrt{4 \pi}} \euler^{\cplx (l + \frac{1}{2}) \phi} \begin{pmatrix} \cplx \Psi_{n,l}(\theta) \\ \pm \Phi_{n,l}(\theta) \end{pmatrix},
\end{align}
where $n \in \N_{0}$, $l \in \{ 0 , \ldots , n \}$, and
\begin{align}
 &\Phi_{n,l}(\theta) = \cos^{l+1} \frac{\theta}{2} \, \sin^{l} \frac{\theta}{2} \, P_{n-l}^{(l,l+1)}(\cos \theta),
\\
 &\Psi_{n,l}(\theta) = (-1)^{n-l} \Phi_{n,l} (\pi - \theta),
\\
 &c_{2}(n,l) = \frac{\sqrt{(n-l)! \, (n + l + 1)!}}{n!},
\end{align}
with the Jacobi polynomials $P_{n}^{(\alpha , \beta)}$.
The following properties of the eigenfunction deserve particular attention:
\begin{align}
 \psi^{(s)}_{\pm, n, l} (\theta, \phi + 2 \pi) &= - \psi^{(s)}_{\pm, n, l}(\theta , \phi),\label{eq:720}
\\
 \psi^{(s)}_{\pm, n, l=0} (\theta = 0 , \phi) &= \sqrt{\frac{n + 1}{16 \pi} } \, \euler^{s \cplx \frac{\phi}{2}} \begin{pmatrix} 1 - s \\ \pm (1 + s) \end{pmatrix} \! ,
\label{eq:np}
\\
 \psi^{(s)}_{\pm, n, l=0} (\theta = \pi , \phi) &= \cplx  (-1)^{n} \sqrt{\frac{n + 1}{16 \pi}} \euler^{s \cplx \frac{\phi}{2}} \begin{pmatrix} 1 + s \\ \pm (1 - s) \end{pmatrix} \! \text{.}
\label{eq:sp}
\end{align}
Equation \eqref{eq:720} shows that the eigenspinors pick up a minus
sign upon a $2\pi$ rotation. Equations~(\ref{eq:np}, \ref{eq:sp})
reveal that the eigenspinors are not well defined at the poles, as an
ambiguous $\phi$ dependence remains. This is similar to the residual 
$\phi$ dependence of the vielbein at the poles.

Let us now study these properties with the global spin base
constructed above.  The Dirac matrices $\gamma_{\mu}$ of
\Eqref{eq:curved_gamma_polar_coordinates} and
${\gamma_{(\vielbein)}}_{\mu}$ are connected via the spin-base
transformation \eqref{eq:sbtrafo},  
 $\gamma_{\mu} = \mcS {\gamma_{(\vielbein)}}_{\mu} \mcS^{-1}$.
%
The corresponding spin connection $\Gamma_{\mu}$ can be calculated from
\begin{align}
 \Gamma_{\mu} = \mcS {\Gamma_{(\vielbein)}}_{\mu} \mcS^{-1} - ( \partial_{\mu} \mcS ) \mcS^{-1},
\end{align}
leading to
 $\Gamma_{\mu} = \frac{\cplx}{2} \gamma_{\mu}$.
The eigenfunctions $\hat{\psi}^{(s)}_{\pm, n, l}$ of the Dirac operator $\slashed{\nabla} = \gamma^{\mu} (\partial_{\mu} + \Gamma_{\mu})$ in the global spin base are then given by
\begin{align}
 \hat{\psi}^{(s)}_{\pm, n, l}(\theta, \phi) = \mcS(\theta, \phi) \psi^{(s)}_{\pm, n, l}(\theta, \phi) \text{.}
\end{align}
It is now straightforward to check that these eigenfunctions are globally well behaved, in particular
\begin{align}
 \hat{\psi}^{(s)}_{\pm, n, l}(\theta, \phi + 2 \pi) &= \hat{\psi}^{(s)}_{\pm, n, l}(\theta, \phi),
\\
 \hat{\psi}^{(s)}_{\pm, n, l=0}(\theta = 0, \phi) &= \cplx \sqrt{\frac{n + 1}{16 \pi}} \begin{pmatrix} \pm (1 + s) \\ 1 - s \end{pmatrix} \! ,
\\
 \hat{\psi}^{(s)}_{\pm, n, l=0}(\theta = \pi, \phi) &= \cplx (-1)^{n} \sqrt{\frac{n + 1}{16 \pi}} \begin{pmatrix} 1 + s \\ \pm(1 - s) \end{pmatrix} \! \text{.}
\end{align}
Not only has the ambiguous $\phi$ dependence disappeared at the poles,
but also the spinors have become $2\pi$-periodic. Since the
eigenfunctions form a complete set of spinor functions on the
2-sphere, we have found a spin base that permits to span functions
on the sphere in terms of globally defined smooth base spinors. This
can serve as a convenient starting point for the construction of
functional integrals for quantized fermion fields.

\section{Conclusion}
\label{sec:conc}

We have emphasized the importance of spin-base invariance for the
description of fermionic degrees of freedom in relativistic
theories. Local SL($d_\gamma,\C$) spin-base invariance is a (hidden)
symmetry of relativistic theories without adding any new propagating
gauge degrees of freedom to the theory in flat space. In curved space,
the associated dynamical degrees of freedom exactly correspond to
those of general relativity. Whereas general covariance and spin-base
invariance seem hardwired to each other via the Clifford algebra, we
have stressed in this work that the associated symmetry
transformations can be used fully independent of each other.

We have demonstrated this explicitly by constructing a global spin base
on a 2-sphere which does not permit an equally globally well-defined
choice of space coordinates. In other words, the coordinate patches
required to cover a manifold do not have to be in one-to-one
correspondence with the spin-base patches that cover the spinor space
at all points of the manifold. 

We consider this mutual independence of general covariance and
spin-base invariance as an indication for the fact that the metric
should not be viewed as more fundamental than the spin structure or
vice versa. Both symmetries should therefore be a direct or emergent
property of a more fundamental theory for matter and gravity. 

In \cite{Gies:2013noa}, we have already shown that quantum gravity
theories that quantize the metric (e.g., in terms of a functional
integral over $g_{\mu\nu}$) preserve spin-base invariance. This
includes, for instance, the asymptotic safety scenario
\cite{Weinberg:1980gg} for quantum Einstein gravity
\cite{Reuter:1996cp}. The mutual independence of general covariance
and spin-base invariance is also a reason for the fact that a
quantization of the spin-base degrees of freedom is not necessary
\cite{Gies:2013noa}, though certainly possible and legitimate
\cite{Harst:2012ni,Dona:2012am}.

\acknowledgments 

The authors thank Martin Ammon, Martin Reuter, Ren\'{e} Sondenheimer,
and Andreas Wipf for valuable discussions. We acknowledge support by
the DFG under grants Gi~328/5-2 (Heisenberg program), GRK1523 and FOR
723.



\begin{thebibliography}{89}

\bibitem{Ashtekar:2014kba} 
  A.~Ashtekar, M.~Reuter and C.~Rovelli,
  arXiv:1408.4336.


\bibitem{Percacci:2002ie} 
  R.~Percacci and D.~Perini,
  Phys.\ Rev.\ D {\bf 67}, 081503 (2003)
  [hep-th/0207033]; 
%
  Phys.\ Rev.\ D {\bf 68}, 044018 (2003)
  [hep-th/0304222].


\bibitem{Eichhorn:2011pc} 
  A.~Eichhorn and H.~Gies,
  New J.\ Phys.\  {\bf 13}, 125012 (2011)
  [arXiv:1104.5366]; 
%
  A.~Eichhorn,
  J.\ Phys.\ Conf.\ Ser.\  {\bf 360}, 012057 (2012)
  [arXiv:1109.3784].


\bibitem{Dona:2013qba} 
  P.~Donà, A.~Eichhorn and R.~Percacci,
  Phys.\ Rev.\ D {\bf 89}, 084035 (2014)
  [arXiv:1311.2898]; 
%
  arXiv:1410.4411.


\bibitem{Dona:2012am} 
  P.~Donà and R.~Percacci,
  Phys.\ Rev.\ D {\bf 87}, no. 4, 045002 (2013)
  [arXiv:1209.3649].


\bibitem{Eichhorn:2015bna} 
  A.~Eichhorn,
  arXiv:1501.05848.


\bibitem{Kraus:2002sa} 
  P.~Kraus and E.~T.~Tomboulis,
  Phys.\ Rev.\ D {\bf 66}, 045015 (2002)
  [hep-th/0203221].


\bibitem{Hebecker:2003iw} 
  A.~Hebecker and C.~Wetterich,
  Phys.\ Lett.\ B {\bf 574}, 269 (2003)
  [hep-th/0307109]; 
  C.~Wetterich,
  Phys.\ Rev.\ D {\bf 70}, 105004 (2004)
  [hep-th/0307145].


\bibitem{Diakonov:2011im} 
  D.~Diakonov,
  arXiv:1109.0091; 
  A.~A.~Vladimirov and D.~Diakonov,
  Phys.\ Rev.\ D {\bf 86}, 104019 (2012)
  [arXiv:1208.1254].


\bibitem{Weyl:1929}
  H.~Weyl, 
  Z.\ Phys.\ {\bf 56}, 330 (1929); 
%
  V.~Fock and D.~Ivanenko,
  Compt.\ Rend.\ Acad.\ Sci.\ Paris {\bf 188}, 1470, (1929); 
%
  B.~S.~DeWitt,
  ``Dynamical theory of groups and fields,''
  Gordon \& Breach, New York, 1965; 
%
  I.~L.~Buchbinder, S.~D.~Odintsov and I.~L.~Shapiro,
  ``Effective action in quantum gravity,''
  Bristol, UK: IOP (1992).


\bibitem{Schroedinger:1932}
  E.~Schr\"odinger, 
  Sitz.ber.~Preuss.~Akad.~Wiss.~(Berlin), Phys.-math.~Kl., 105 (1932).


\bibitem{Bargmann:1932}
  V.~Bargmann, 
  Sitz.ber.~Preuss.~Akad.~Wiss.~(Berlin), Phys.-math.~Kl., 346 (1932).


\bibitem{Kofink:1949}
  W.~Kofink,
  Math.\ Z.\ {\bf 51}, 702 (1949).


\bibitem{Brill:1957fx} 
  D.~R.~Brill and J.~A.~Wheeler,
  Rev.\ Mod.\ Phys.\  {\bf 29}, 465 (1957).


\bibitem{Unruh:1974bw} 
  W.~G.~Unruh,
  Phys.\ Rev.\ D {\bf 10}, 3194 (1974).


\bibitem{Finster:1998ws} 
  F.~Finster, J.~Smoller and S.~-T.~Yau,
  Phys.\ Rev.\ D {\bf 59}, 104020 (1999)
  [gr-qc/9801079].


\bibitem{Casals:2012es} 
  M.~Casals, S.~R.~Dolan, B.~C.~Nolan, A.~C.~Ottewill and E.~Winstanley,
  Phys.\  Rev.\ D {\bf 87}, 064027 (2013)
  [arXiv:1207.7089].


\bibitem{Gies:2013dca}
  H.~Gies and S.~Lippoldt,
  Phys.\ Rev.\ D {\bf 87}, 104026 (2013)
  [arXiv:1303.4253].

\bibitem{Christiansen:2015}
N.~Christiansen, K.~Falls, J.~Meibohm, J.~M.~Pawlowski, 
M.~Reichert, in preparation.

\bibitem{Weldon:2000fr} 
  H.~A.~Weldon,
  Phys.\ Rev.\ D {\bf 63}, 104010 (2001)
  [gr-qc/0009086].


\bibitem{Gies:2013noa} 
  H.~Gies and S.~Lippoldt,
  Phys.\ Rev.\ D {\bf 89}, 064040 (2014)
  [arXiv:1310.2509].


\bibitem{Pauli:1936}
  W.~Pauli,
  Ann.\ Inst.\ Henri Poincar\'{e} {\bf 6}, 109 (1936).


\bibitem{Lippoldt:2012}
  S.~Lippoldt,
  master thesis, Jena (2012).


\bibitem{Lippoldt:2015}
  S.~Lippoldt, in preparation.


\bibitem{Lawson:1990}
  H.~B.~Lawson and M.-L.~Michelsohn, 
  Princeton Math. Ser. {\bf 38}, Princeton University Press (1990).


\bibitem{Cornwell:1989bx} 
  J.~F.~Cornwell,
  ``Group Theory In Physics. Vol. 3: Supersymmetries And Infinite Dimensional Algebras,''
  London, UK: Academic (1989) 628 p. (Techniques of physics, 10).


\bibitem{Ogievetsky:1965ii} 
  V.~I.~Ogievetsky and I.~V.~Polubarinov,
  Sov.\ Phys.\ JETP {\bf 21}, 1093 (1965)
  [Zh.\ Eksp.\ Teor.\ Fiz.\  {\bf 48}, 1625 (1965)].


\bibitem{Woodard:1984sj} 
  R.~P.~Woodard,
  Phys.\ Lett.\ B {\bf 148}, 440 (1984).


\bibitem{Pitts:2011jv} 
  J.~B.~Pitts,
  Stud.\ Hist.\ Philos.\ Mod.\ Phys.\  {\bf 43}, 1 (2012)
  [arXiv:1111.4586].


\bibitem{Camporesi:1995fb} 
  R.~Camporesi and A.~Higuchi,
  J.\ Geom.\ Phys.\  {\bf 20}, 1 (1996)
  [gr-qc/9505009].


\bibitem{Weinberg:1980gg}
  S.~Weinberg,
{\it  In *Hawking, S.W., Israel, W.: General Relativity*, 790-831}.


\bibitem{Reuter:1996cp}
  M.~Reuter,
  Phys.\ Rev.\  D {\bf 57}, 971 (1998)
  [arXiv:hep-th/9605030]; 
  M.~Niedermaier and M.~Reuter,
  Living Rev.\ Rel.\  {\bf 9}, 5 (2006); 
  R.~Percacci,
  In *Oriti, D. (ed.): Approaches to quantum gravity* 111-128
  [arXiv:0709.3851]; 
  M.~Reuter and F.~Saueressig,
  New J.\ Phys.\  {\bf 14}, 055022 (2012)
  [arXiv:1202.2274].


\bibitem{Harst:2012ni} 
  U.~Harst and M.~Reuter,
  JHEP {\bf 1205}, 005 (2012)
  [arXiv:1203.2158].


\end{thebibliography}
\end{document}